\documentclass[12pt,a4paper]{article} 
\usepackage{graphics}
 \usepackage{afterpage}
 \usepackage{enumerate}
\usepackage[pdftex]{graphicx}

\begin{document}

 \title{\bf Total ozone columns and vertical ozone profiles above Kiev in 2005-2008}

  \author{
    A.V. Shavrina$^1$, 
    M. Kroon$^2$,
    V.A. Sheminova$^1$,
    Ya.V. Pavlenko$^1$,\\
    A.A. Veles$^1$,
    I.I. Synyavski$^1$,
    Ya.O. Romanyuk$^1$,
    }
\date{}
 \maketitle
 \begin{center}{
          $^1$Main Astronomical Observatory,
          National Academy of Sciences of the Ukraine,\\
           27 Akademika Zabolotnoho St.,
           03680 Kiev, Ukraine

           $^2$Royal Netherlands Meteorological
           Institute (KNMI), The Netherlands}
\end{center}

\begin{abstract}

The study of total ozone columns above Kiev
and variations of ozone concentrations in the troposphere at different
altitudes above Kiev was carried out using ground-based
Fourier Transform
InfraRed (FTIR) spectrometric observations
that are taken on a routine basis at the Main Astronomical Observatory
of the National Academy of Sciences of Ukraine (MAO NASU). This study
was performed within the framework of the
international
ESA-NIVR-KNMI OMI-AO project no.~2907 entitled
OMI validation by ground-based remote sensing: ozone
columns and atmospheric profiles during the time
frame 2005--2008. The infrared FTIR spectral observations of direct
solar radiation in the wavelength range of 2-12 micron as transmitted
through the Earth's atmosphere were performed during
the months of April-October of each year. The aim of the project was
the validation of total ozone columns and vertical ozone profiles as
obtained by the Ozone Monitoring Instrument (OMI)) onboard of the NASA
EOS-Aura scientific satellite platform. The modeling of the ozone spectral band
shape near 9.6 microns was performing with the MODTRAN code and a
molecular band model based on the HITRAN-2004 molecular database. The
a-priori information for the spectral modeling consisted of water vapor
and temperature profiles from the NASA EOS-Aqua-AIRS satellite instrument,
stratospheric ozone profiles from the NASA EOS-Aura-MLS satellite
instrument, TEMIS-KNMI climatological ozone profiles and surface ozone 
concentration measurements performed at the specific times of infrared spectra 
observations.
The troposphere ozone variability was
analyzed for two typical episodes: the spring episode of enhanced total
ozone columns and the summer episode of enhanced surface ozone
concentrations.

\end{abstract}

\section{Introduction}
The study of ozone and its
variability in the Earth's atmosphere is of critical
importance to both the scientific world, to international policy makers
and to the general public at large. The stratospheric ozone layer
protects all living organisms on Earth from the harmful effects of
excessive ultraviolet solar irradiation.

Close monitoring of the recovery of the
global ozone layer as a result of the Montreal protocol and its
amendments on limiting the release of ozone depleting substances has
therefore become a subject of major attention in atmospheric sciences
and high levels in society.

However, the growing
presence of high concentrations of ozone in the lower part of the
atmosphere, the troposphere, where it acts as a health hazardous air
pollutant and also as an important greenhouse gas is an alarming and
disturbing development. Tropospheric ozone is very toxic to air
breathing organisms like humans and corrosive to the leaf surfaces of
plants. This breathing-level ozone is formed by photochemical reactions
of gaseous precursors of natural origin such as volatile organic
compounds released by trees but also by anthropogenic sources such as
exhaust emissions from motor vehicles and industrial enterprises. The
observation of high concentrations of surface ozone over urbanized and
industrialized regions on our blue marble planet is a clear indicator
of substantial anthropogenic pollution of the air we breathe.

To identify the key processes in the stratospheric ozone budget,
satellite soundings of the atmospheric abundance of the relevant trace
gases have been routinely performed since the late
1970's. These satellite observations allow us to
determine the total ozone column and vertical profiles of ozone in the
atmosphere, from the stratosphere down into the troposphere. Here
ground-based observations serve the purpose of confirming the satellite
data by means of validation and also complementing the satellite data
by means of extending the tropospheric observations down to the surface
where most satellite instruments have a reduced sensitivity. This is
especially important to understand the role of local and regional
sources and sinks of tropospheric ozone and its precursors. Such
ground-based observations aid in studying the dynamic behavior of air
pollutants and to verify the models describing the cross-boundary and
intercontinental transport of air pollutants. These studies will also
support the development of national and international policies aimed at
reducing the precursors of tropospheric ozone and particularly
breathing-level ozone, and reducing the effect of tropospheric ozone
acting as a greenhouse gas on local and regional scales.

In our previous publications
Shavrina et al.,
2007,
Shavrina et al.,
2008 the results of our observations on
total ozone column during the time frame 2005-2007 and the retrieval of
tropospheric ozone profiles over Kiev in 2007 was described . This work
was further expanded and improved in 2008 and here we summarize the
results of our research performed during the time frame 2005-2008.

\section{Ground based FTIR ozone observations}

Ground-based observations
of strong and weak absorption lines in the electromagnetic spectrum of
direct solar radiation transmitted through the Earth's
atmosphere to the surface are routinely made at the
Main Astronomical Observatory of National
Academy of Sciences Ukraine
with the Fourier
Transform InfraRed spectrometer (FTIR model  Infralum FT
801, [Egevskaya, Vlasov, Bublikov, 2001]). This particular instrument was 
optically redesigned for the study
of atmosphere pollution in the Kiev metropolitan area [Shavrina et al.,
2007]. The working spectral range of the Fourier spectrometer is 2-12
microns (equivalent to 800-5000
$cm^{-1}$ with a best spectral resolution of \~{}1.0$cm^{-1}$.
Recording software for the acquisition of the spectra allows us to
average 2-99 individual spectra during one observation period. Here
we averaged over 4
single spectra which spans a 120-180 seconds time frame as was
recommended by the industrial developers of the spectrometer to avoid
the effects of atmospheric instability at longer exposures.
The averaged spectra
have signal to noise ratio S/N = 150-200 which is more than
sufficient for our purposes.

\section{Satelite based ozone observations}

The Ozone Monitoring
Instrument OMI [Levelt et al., 2006] is a contribution of the
Netherlands's Agency for Aerospace Programs (NIVR) in
collaboration with the Finnish Meteorological Institute (FMI) to the
NASA EOS-Aura satellite platform [Schoeberl et al., 2006] launched in
July 2004. The OMI instrument currently continues the global total
ozone column measurements recorded by the NASA Total Ozone Mapping
Spectrometer (TOMS) instruments since 1978. A series of TOMS
instruments was operative on the Nimbus-7 (1978-1993) and Earth Probe
(EP) (1996-2005) satellite platforms. The OMI instrument is a nadir
viewing wide swath UVVIS hyperspectral spectrometer measuring solar
light reflected and backscattered from the Earth's
atmosphere and surface in the wavelength range from 270 nm to 500 nm
with a spectral resolution of 0.45 nm in the ultraviolet and 0.63 nm in
the visible. The instrument has a 2600 km wide viewing swath and is
capable of daily, global contiguous mapping of various atmospheric
trace gases, clouds, surface UV radiance levels and particulate
atmospheric species with an unprecedented high spatial resolution of
13x24 $km^{2}$ at nadir. The OMI total ozone column data used in this work
were obtained from the OMI-TOMS algorithm which is based on the
long-standing NASA TOMS V8 retrieval algorithm [Bhartia and Wellemeyer,
2002] and the OMI-DOAS algorithm developed at Royal Dutch
Meteorological Institute (KNMI) [Veefkind et al., 2006] which is based
on the Differential Optical Absorption Spectroscopy (DOAS) technique. A
detailed analysis of the similarities and differences between OMI-TOMS
and OMI-DOAS total ozone column data can be found in the recent work of
Kroon et al. [2008].
Here both OMI ozone products are obtained in the new version of the OMI
level 1 (radiance and irradiance) and level 2 (atmospheric data
products) data set named collection 3. Please visit the NASA DISC at
http://disc.gsfc.nasa.gov/Aura/OMI/ for EOS Aura OMI level 2 orbit
data. Please visit the Aura Validation data Center at
http://avdc.gsfc.nasa.gov for EOS-Aura OMI station overpass data.
Please consult the OMI README files for the latest OMI data product
information.

\section{Spectral modeling and analysis}
The estimates of total
ozone columns (TOC) in the Earths atmosphere were obtained using a
simulation of the ozone absorption spectrum at 9.6 microns with the
radiative transfer code MODTRAN4.3 [Bernstein et al., 1996]. The
program calculates the transmission of electromagnetic radiation of
solar origin through the modeled atmosphere and its subsequent
reflection on the surface in the frequency range from 0 to 50000
cm$^{-1}$.
The program employs a two-parameter (temperature and pressure) model of
molecular spectral absorption, which is calculated with the molecular
database HITRAN
(http://www.cfa.harvard.edu/hitran/)
containing molecular spectral absorption lines. To calculate the band
model, data for 12 light gaseous molecules (H2O, CO2, O3, CO, CH4, O2,
NO, SO2, NO2, N2O, NH4 and HNO3) from the HITRAN data base were used,
and for heavy molecules - CFC (9 molecules) and CLONO2 , HNO4, CCl4 and
N2O5 the calculated absorption cross section (see [Bernstein et al.,
1996]) were applied. With the program MODTRAN4.3 the spectrum of solar
radiation is calculated that is to be received by the FTIR instrument
on the ground depending on its viewing geometry. The calculations were
performed in the approximation of local thermodynamic equilibrium (LTE)
for the moderate spectral resolution of 2
cm$^{-1}$
which corresponds almost exactly to our observed spectra. The model
parameters have been calculated by us using the database HITRAN-2004
[Rothman et al., 2005].

To construct the a-priori
atmospheric profiles of ozone, temperature and water vapor we employed
here (i) the {observation of
} surface ozone
concentrations as measured by the ground-based ultraviolet ozonometer
TEI-49i that is located near to the Fourier spectrometer detecting
breathing-level ozone, (ii) data of the NASA Atmospheric Infrared
Sounder (AIRS) satellite instrument
(http://avdc.gsfc.nasa.gov/Data/AIRS/)
flying on board of the NASA EOS-Aqua satellite, and data of the
Microwave Limb Sounder (MLS) satellite instrument
({http://avdc.gsfc.nasa.gov/Data/Aura/})
flying on board of the NASA EOS-Aura satellite. For the analysis of our
observations of 2007-2008, we used MLS data version v2.2, the higher
accuracy of which has enabled us to develop a new method of ozone
profile analysis:
we are now varying the
input profile of the tropospheric ozone only, where we only scale the
stratospheric profile by a
certain factor within
the specified accuracy of MLS (2-5\% in the layers of the pressure
216-0.02 hPa) without modifying its overall shape. A detailed
description of our method for determining detailed tropospheric and
stratospheric concentrations of ozone can be found in [Shavrina et al.,
2007, Shavrina et al., 2008]. The result of the analysis is the best
fit of the model spectrum to the observed spectra of the ozone band,
from which we obtained the best estimates of the tropospheric vertical
profiles of ozone, the total column amount of ozone in the troposphere
and finally the total column amount of ozone in the
atmosphere.

\section{Results}

The results of the
comparison of the OMI collection 2 satellite and FTIR ground-based
total ozone column observations and first examples of the retrieved
vertical ozone profiles as obtained during the time frame 2005-2007
were already presented in [Shavrina et al., 2007, Shavrina et al.,
2008]. In the summer of 2008 a new version
of all the OMI data became available, named collection 3 and
short-named COL3. In
Figure 1 we show a comparison of our ground-based results for the year
2007 with Aura-OMI data of collection 2 for the two algorithm versions
available, OMI-TOMS and OMI-DOAS, respectively. The OMI collection 2
total ozone column data for the year 2007 is on average very close the
ground-based observations, albeit with substantial standard
deviations. The average difference of
satellite minus ground-based amounts to -0.33 DU and -4.32 DU for
OMI-DOAS and OMI-TOMS collection 2 data products respectively, with
10.66 DU and 10.88 DU standard deviations (1.35 DU and 1.39 DU standard
errors). The OMI
collection 3 total ozone column data for the year 2007 is on average not
so close the ground-based observations, albeit with lesser standard
deviations. The results are shown in Figure 2. The average difference of
satellite minus ground-based amounts to -6.63 DU and -7.54 DU for
OMI-DOAS and OMI-TOMS data products respectively, with a 9.06 DU and
9.00 DU standard deviations (1.31 DU and 1.25 DU standard errors).
Destriping of the OMI data towards collection 3 has substantially
reduced the standard deviations of validation comparisons against
reference data but has substantially increased the bias.

Figure 3 presents a
comparison of our ground-based estimates of total ozone columns with
OMI-DOAS and OMI-TOMS TOC satellite data for the year 2008. The average
difference between satellite and
ground-based
observations is 0.29 DU
and 2.85 DU for the OMI-TOMS and OMI-DOAS data products, respectively,
with standard deviations of about 9 DU. Here the standard errors are
equal to 0.99 and 1.18 DU, respectively (see Table 1).
The correlation
coefficient is 0.97 in both cases.
In table 1 we present a
comparison of our results with OMI for the entire observation period
2005-2008, which shows that the obtained accuracy meets the
requirements for validation (3.0\%, see [Shavrina et al.,
2007]). More details can be found in our
before mentioned publications.

\section{Tropospheric ozone profiles }

In a previous publication
[Shavrina et al., 2008] we examined the variability of tropospheric
ozone profiles for carefully selected episodes of 2007, each of which
demonstrated a characteristic and different ozone situation. Two
separate spring episodes showed an increased total and tropospheric
ozone abundance, one of which episodes probably represents the
occurrence of a stratospheric intrusion, i.e., the intrusion of
stratospheric air masses with a much higher concentration of ozone
dwelling down into the troposphere. The following summer episode
demonstrated high concentrations of ground-level ozone probably due to
the photochemical production of ozone, and a subsequent autumn episode
that showed a much lower total ozone amount over Kiev of only 260.2 DU
which is getting pretty close to the upper limit of the so-called
ozone hole (220 DU). For the purpose of
this study presented in this paper we have closely analyzed the changes
of daytime ozone profiles for a set of selected days during these
episodes.
In Figure 4 we show the
spectra observed during the daytime of the
23$^{rd}$
of April 2007. This figure also shows the best fit of our modeled
spectra to the FTIR spectrum observed at
11h15m local time. On
this particular day, the total ozone column amounted to 411.0\ DU, as
recorded by the FTIR and the tropospheric ozone column amounted to 48
DU [Shavrina et al., 2008], which both were very high. Please note that
the highest value of the total ozone column in 2007 over Kiev was
recorded on April 22$^{nd}$ and measured 448 DU as revealed by the OMI
 satellite data. Figure 5
shows the retrieved vertical profiles of atmospheric ozone for the time
frame of April 23-28, 2007 for the moments of observations with a
maximum solar elevation. Figure 5 reveals strong changes in the ozone
abundance with altitude but contains too little information to reveal
the intricate details of the intrusions process. Subsequently, in
Figure 6 we plot all the recorded profiles as a function of time where
we clearly see the intrusion of the stratospheric ozone rich air into
the troposphere, its sinking during the time frame 23-26 April and a
further dissipation and sinking to almost the surface during the
26$^{th}$-28$^{th}$
of April. Please note the high level of detail presented in this
one-week plot revealing the temporal and vertical spatial resolution
and high accuracy of our ground-based method that can be achieved on a
continuous basis. We believe that the daytime FTIR ozone profiles
clearly reveal the intricate dynamics of the ozone layer caused by the
intrusion of stratospheric ozone rich air to the lower altitude
troposphere layers on April
23$^{rd}$ and its sinking and dissipation over the course of the
 following few days.

Figure 7 shows the
reconstructed profiles of atmospheric ozone for the 18$^{th}$
of July 2007 recorded at 13h35m, 14h52m, 16h10m, 17h10m, 18h15m and
19h27m of local time. The enhanced abundance of ozone in the lower
troposphere close to the surface, the high values of surface ozone
concentrations and
their daily dynamics are typical for the summer episodes of high ozone
concentration in the troposphere due to photochemical processes. Note
that on this day the total ozone column amount is rather low (291.5
DU). Figure 8 shows one day
(at July 18, 2007) in our
summer episode of
continuous data, revealing photochemical production of ozone in the
surface layers of our polluted urban
atmosphere.
Please note the high
level of detail presented in this single day plot revealing the
vertical resolution and accuracy of our ground-based method that can be
achieved on a daily basis.

In the course of 2008 the final operational Aura-OMI data set of atmospheric 
vertical ozone profiles,
labeled OMO3PR, became publically available and we compared them with
our retrieved profiles for validation. A first comparison of the satellite ozone
profile data set to our ground-based ozone profile data set is shown in the 
left hand side image of Figure 9 and
it shows rather large differences between the profiles particularly for the
tropospheric pressure layers. 
The surface ozone concentrations as reported in our ground-based retrieved 
ozone profiles are fixed by the simultaneous in-situ measurements of 
surface ozone concentrations as performed at the MAO NASU with the ozonometer. 
With the OMI  satellite instrument being rather blind to the lowest layers of 
the atmosphere the satellite data retrieval algorithm tends to use the ozone 
concentration in the lowest - say tropospheric - pressure layers as an 
additional degree of freedom to increase the quality of the spectral fit, 
however, leading to large overestimations of the tropospheric ozone 
concentrations. Other validation studies confirmed these large deviations and 
hence the algorithm was redesigned. The latest version of the OMO3PR data set 
became available on 09 September 2009 and this data set complies much better 
with our ground-based data set. As an example, the fit of the latest OMO3PR 
data to our retrieved ozone profile for April 23, 
2007 as shown in the right hand side image of Figure 9 looks very promising.

\section{Conclusion}
During the time frame
2005-2008 we have obtained an elaborate data set of total ozone column
and vertical ozone profile estimates from our ground-based observations
with the optically adapted Fourier Transform InfraRed spectrometer. Our
estimates of total ozone are in a good agreement with the
satellite-based EOS Aura-OMI total ozone columns. The differences are
in the range of a few percent, which meets the validation requirements.
The analysis of our retrieved ozone profiles was performed for two
representative cases of the tropospheric ozone dynamics: a spring
episode of high total ozone and a summer episode of photochemical ozone
production. We noted also a fall episode of low total amount of
stratospheric ozone.
We
firmly believe that the
high vertical and temporal resolution of our
daytime
FTIR observations
enables us to
clearly
observe the intricate dynamics of the ozone layer caused by the
intrusion of stratospheric ozone rich air to the lower altitude
troposphere layers on April
23$^{rd}$
and its sinking and dissipation over the course of the following few
days. We also firmly believe that the daytime FTIR ozone profiles are
accurate enough to clearly show the occurrence of high tropospheric
ozone concetrations near the surface as a results of photochemical
processes in response to increased air
pollution. A first
quantitative comparison of our retrieved vertical ozone profiles with
the data Aura-OMI OMO3PR indicated some significant differences for the
lower troposphere. However, the fit of the latest version of the publically 
available OMO3PR data set to our retrieved ozone profile for April 23 2007 
looks very promising. Please note that the reported values of surface ozone 
concentrations in our retrieved ozone profiles are originating from the 
in-situ measurements of surface ozone concentrations as performed at the MAO 
NASU and hence are fixed by true observations. 
More attention will be given to 
this work in the near future with the complete OMO3PR data set now available. 
However, any airplane or lidar measurements of ozone at different altitudes 
over Kiev, which could be arbitral, are absent at the time of writing.

Our methodology could in principle be used for any FTIR observatinal
station allowimg for a better characterization of the horizontal and vertical 
dynamics of the ozone layer world wide from ground-based observations alone.

\section{Acknowledgements}
The authors thank the
administrations of the AVDC, Aura-MLS, Aura-TES and Aqua-AIRS web sites
for the provision of the necessary satellite data of the respective
atmosphere soundings. The work was partially supported by a grant STCU
(2005-2007) and the Space Agency of Ukraine (2007-2008).


  \begin{figure}
  \centering
 \includegraphics[width=\textwidth]{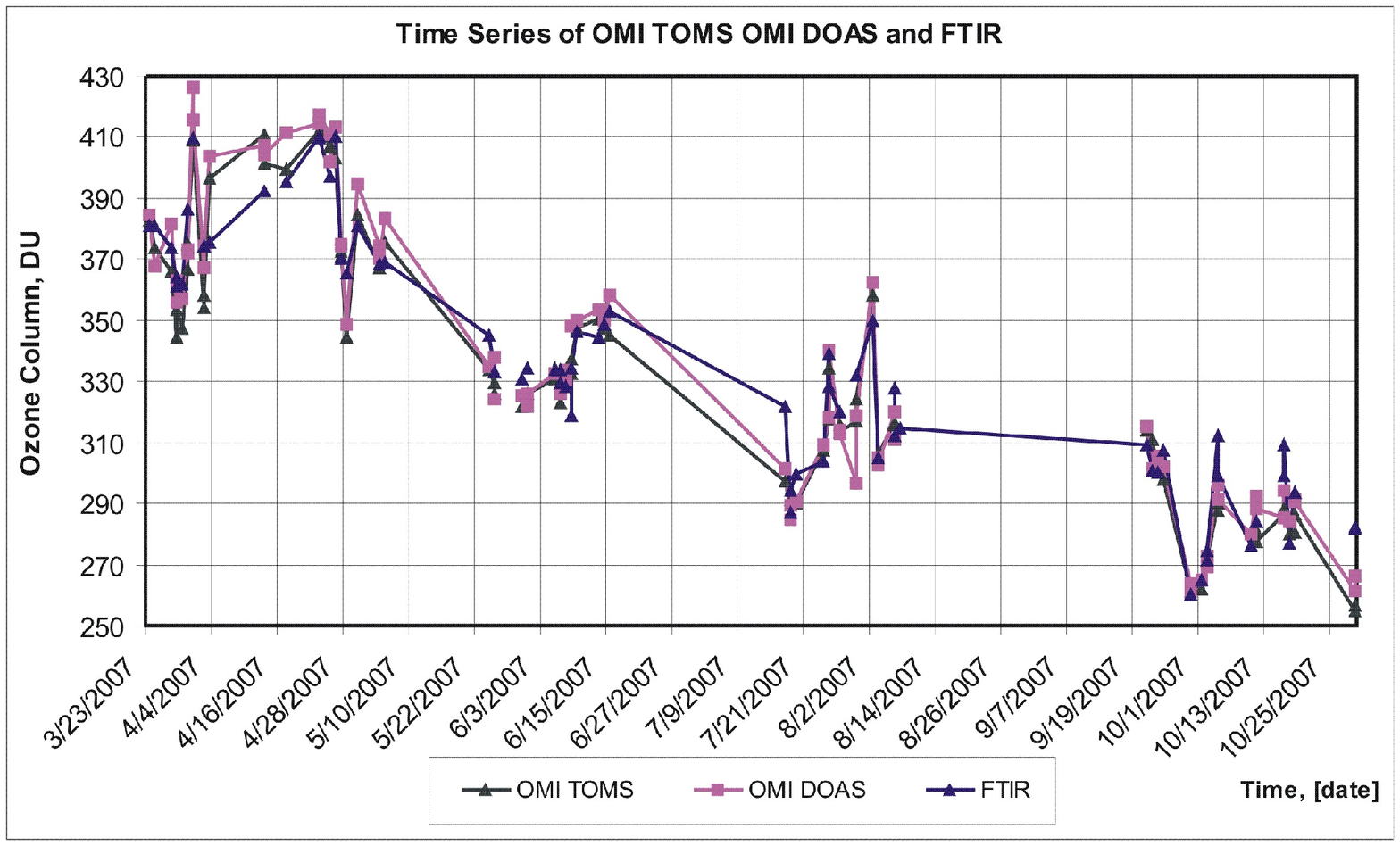}
      \caption{
 Time series of OMI satellite total ozone columns of Collection 2 
and the ground-based FTIR total ozone column of 2007 for Kiev (MAO). The 
average difference of satellite minus ground-based amounts to -0.33 DU and -4.32
DU for OMI-DOAS and OMI-TOMS data products respectively, with 10.66 DU 
and 10.88 DU standard deviations (1.35 DU and 1.39 DU standard errors).      
              }
         \label{Fig1}
   \end{figure}

  \begin{figure}
  \centering
 \includegraphics[width=\textwidth]{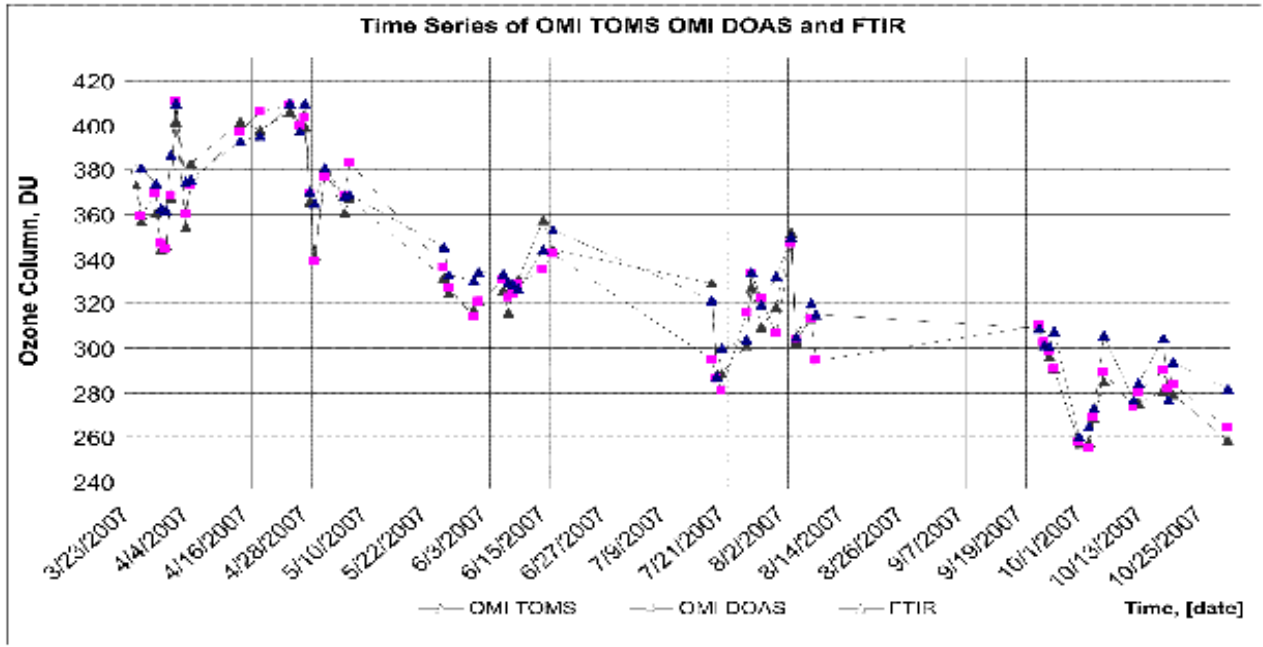}
      \caption{
 Time series of the OMI satellite total ozone columns (Collection 3 
data available as of 2008) and the ground-based FTIR total ozone column of 2007
for Kiev (MAO). The average difference of satellite minus ground-based amounts 
to -6.63 DU and -7.54 DU for OMI-DOAS and OMI-TOMS data products respectively, 
with a 9.06 DU and 9.00 DU standard deviations (1.31 DU and 1.25 DU standard 
errors). The coefficient of correlation is 0.98 for both sequences with total 
number collocated FTIR and OMI-DOAS and OMI-TOMS data of 52 and 54, 
respectively. 
              }
         \label{Fig2}
   \end{figure}

  \begin{figure}
  \centering
 \includegraphics[width=\textwidth]{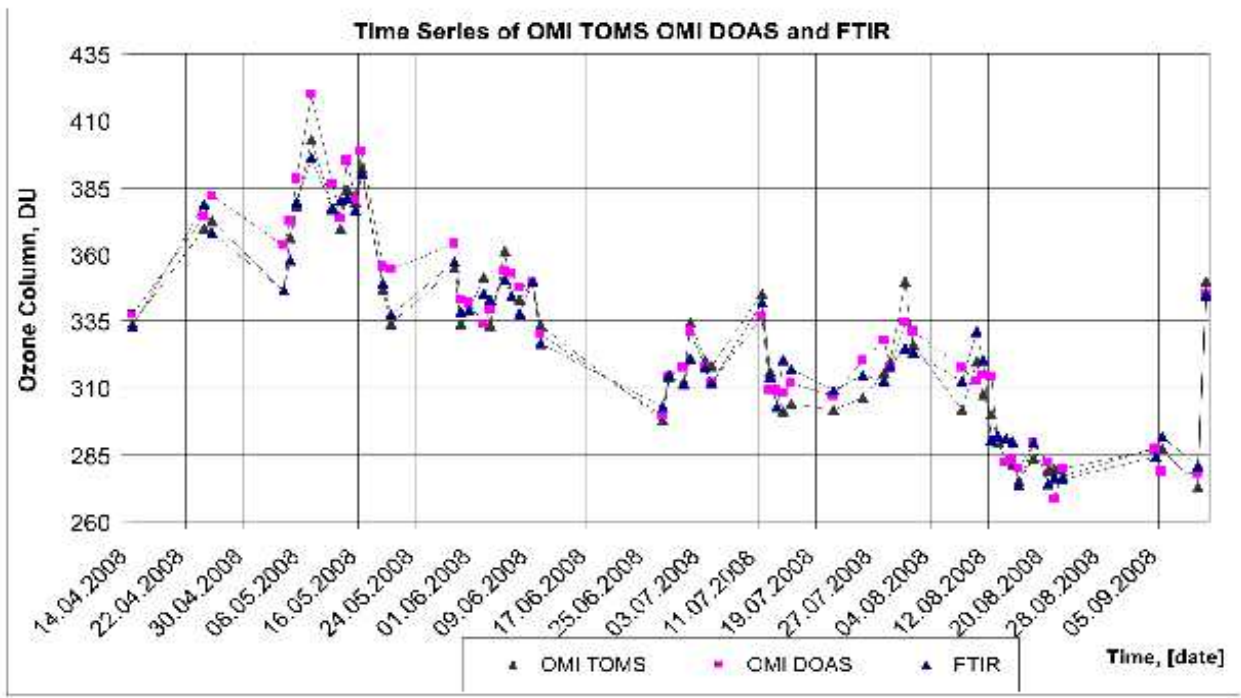}
      \caption{
 Time series of the satellite OMI total ozone column and the 
ground-based FTIR total ozone data of 2008 for Kiev (MAO). Average difference 
of satellite minus ground-based amounts to 2.85 DU and -0.29 DU for OMI-DOAS 
and OMI-TOMS respectively, with a 8.88 DU and 7.52 DU standard deviations 
(1.18 DU and 0.99 DU standard errors). Coefficient of correlation is 0.97 for 
both sequences with total number collocated FTIR and OMI-DOAS and OMI-TOMS data 
of 57. 
              }
         \label{Fig3}
   \end{figure}

  \begin{figure}  \centering
 \includegraphics[width=8.cm]{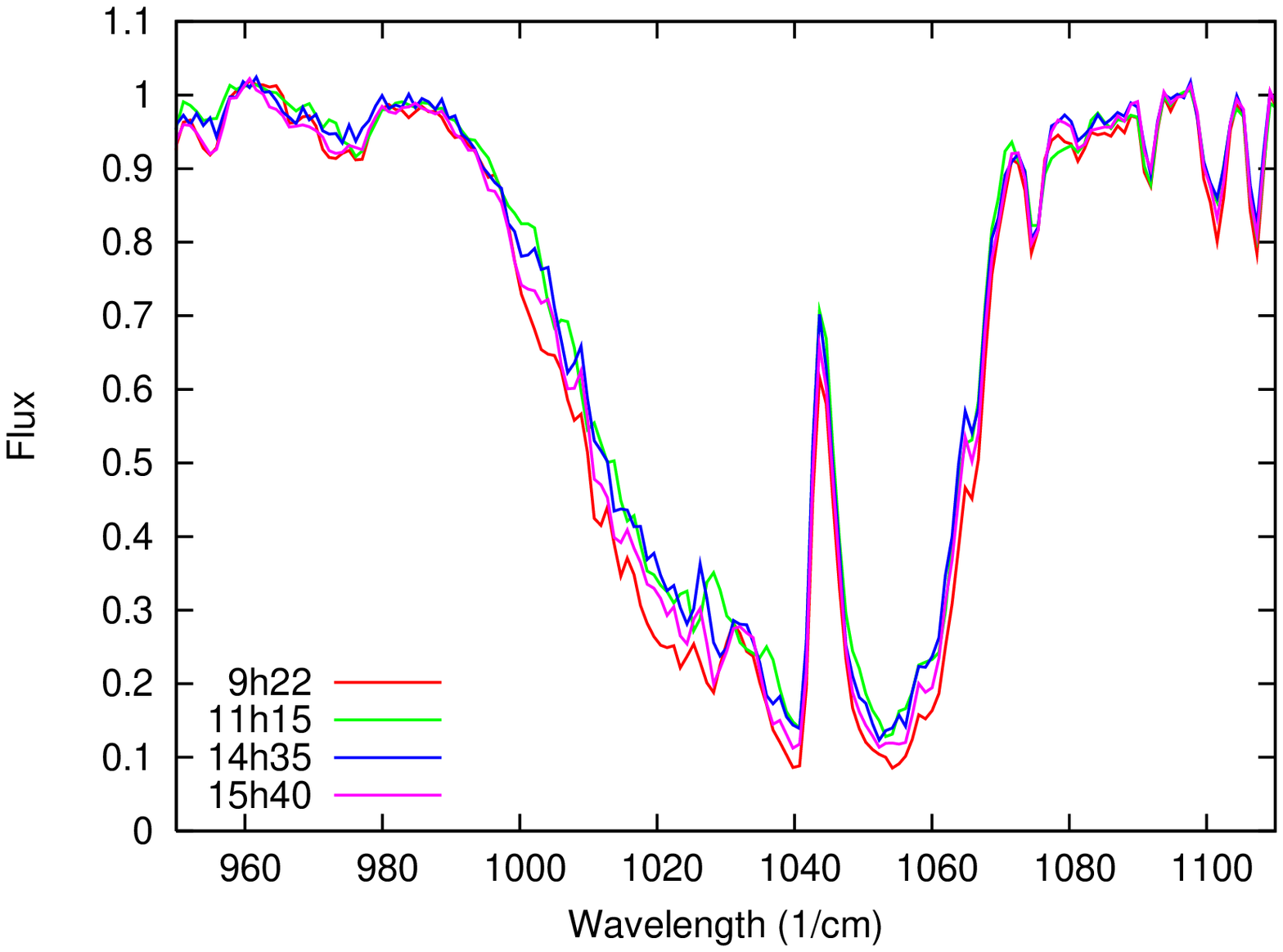}
 \includegraphics[width=8.cm]{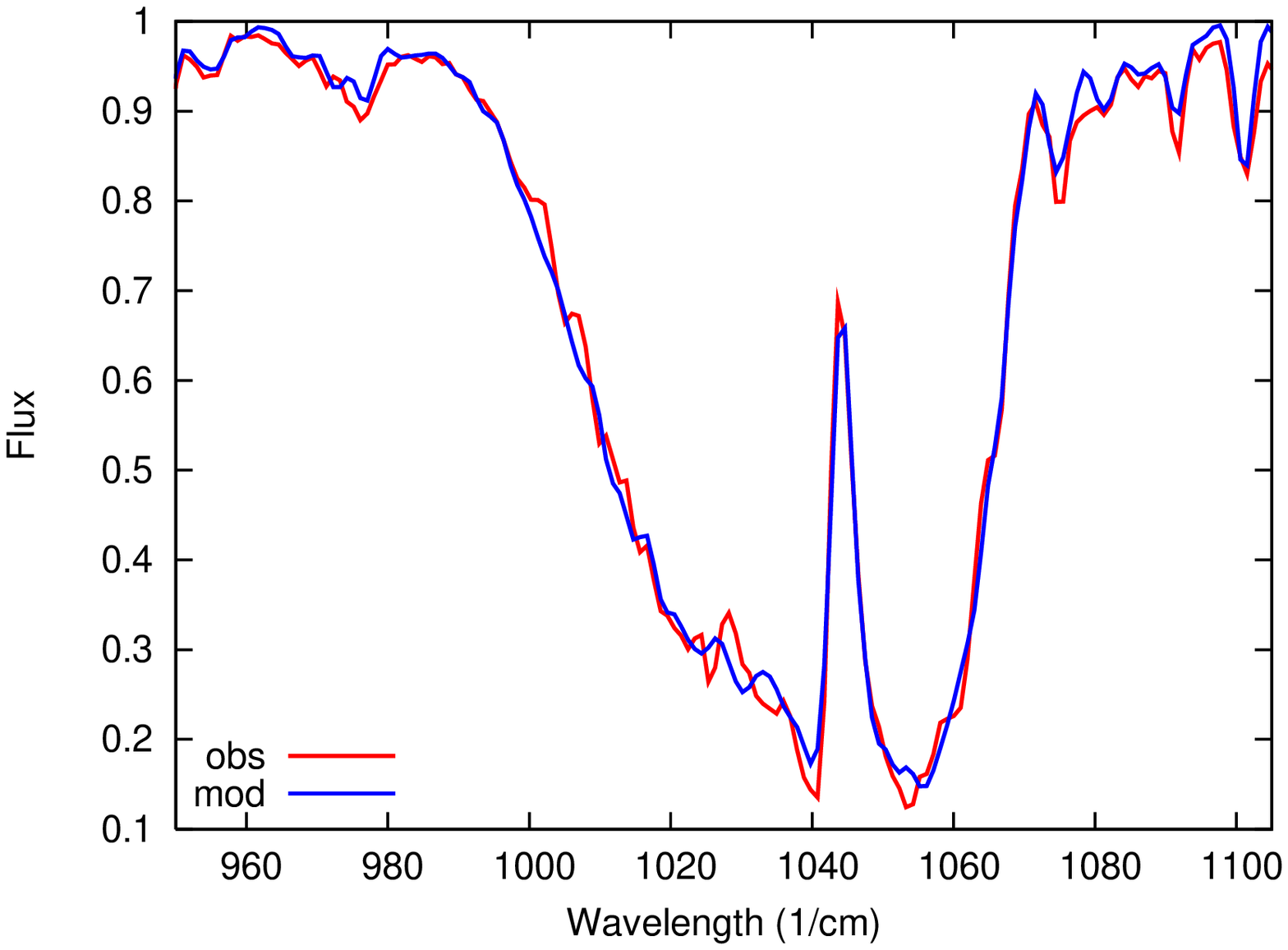}
      \caption{
The spectra observed during the 23rd of April 2007. (r.h.s.) and
the best fit of the modeled spectra to the recorded FTIR spectrum observed at 
11h 15 min local time.
              }
         \label{Fig4}
   \end{figure}

  \begin{figure}
  \centering
\includegraphics[width=8cm]{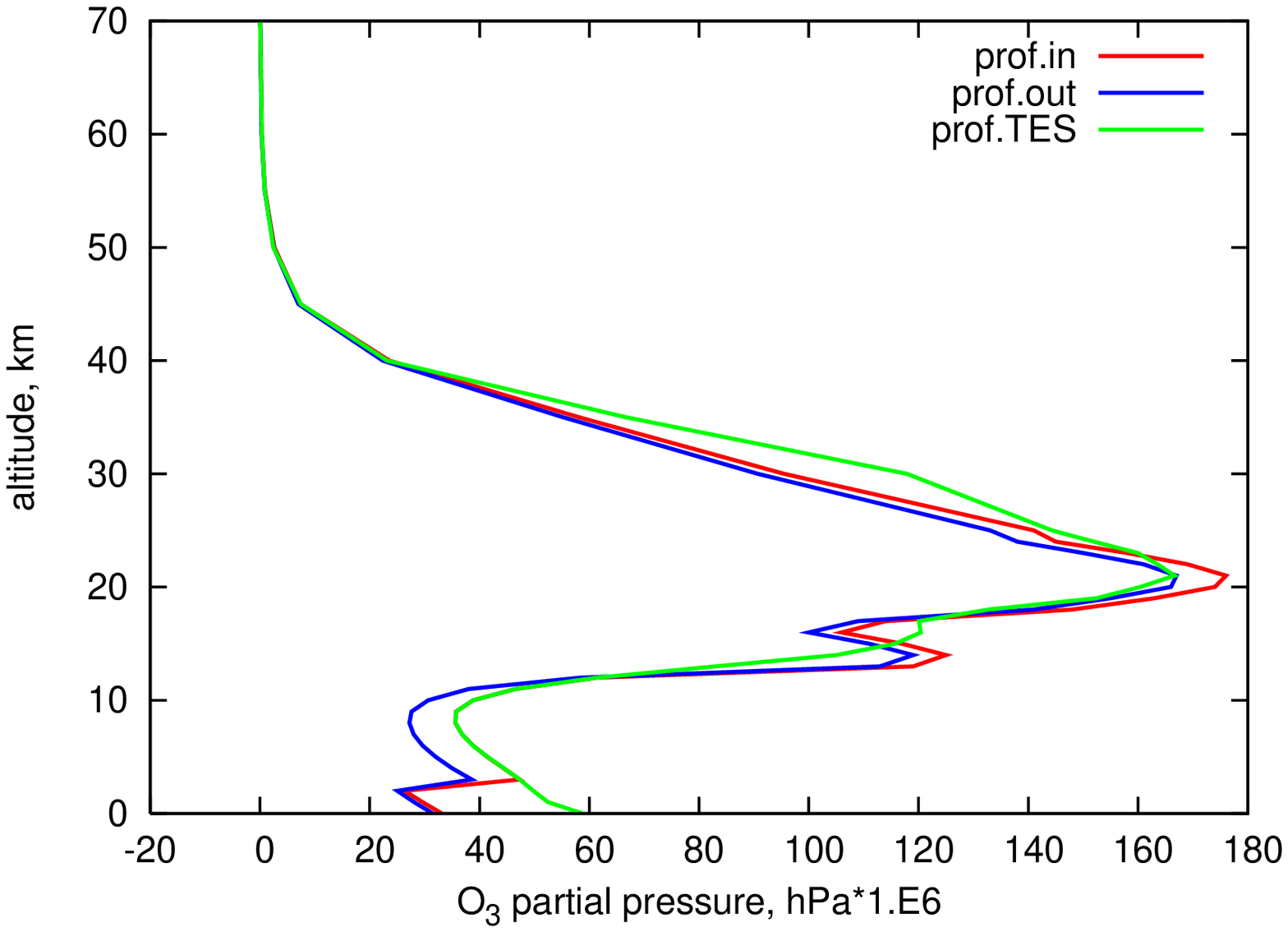}
\includegraphics[width=8cm]{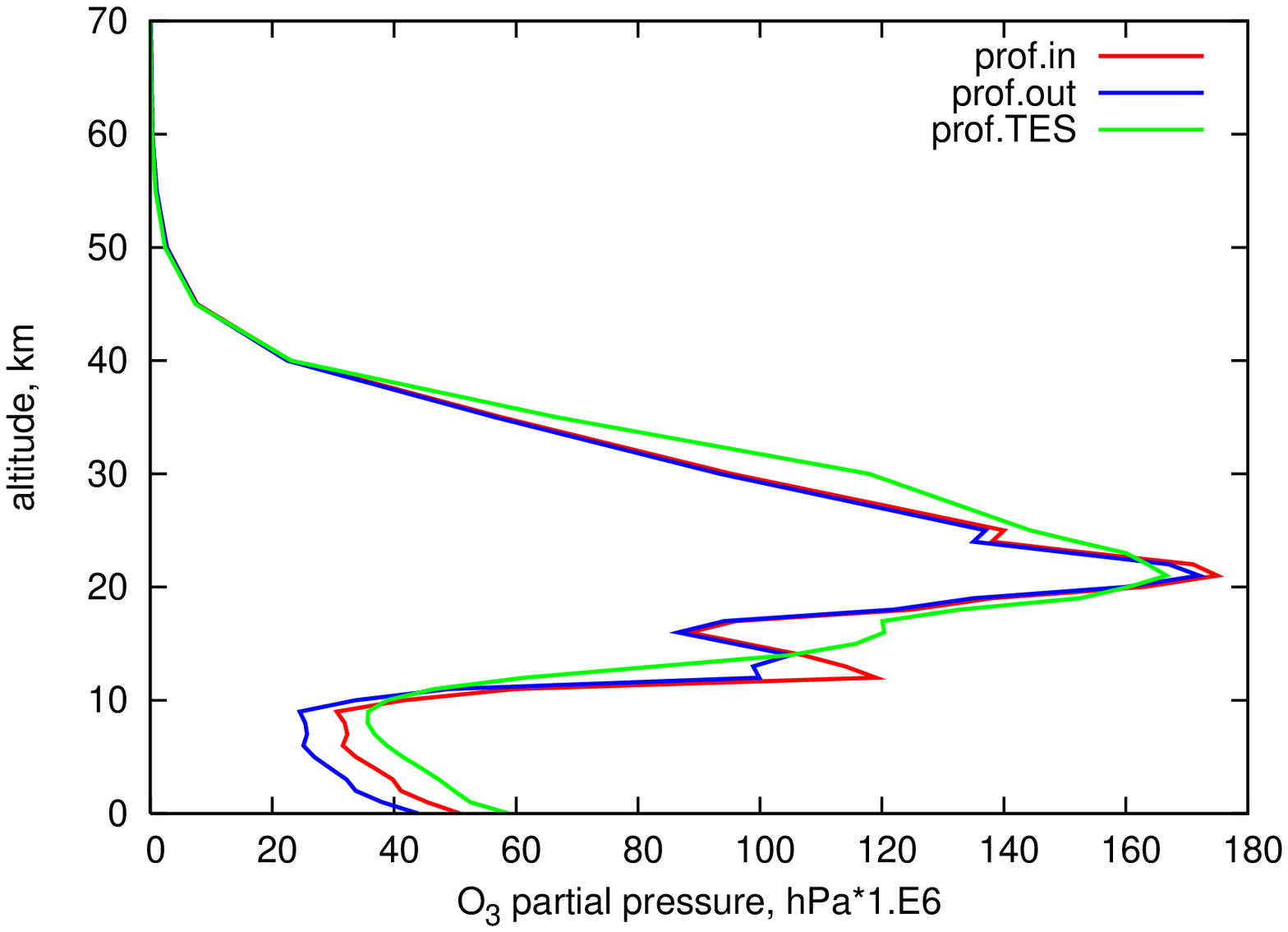}
\includegraphics[width=8cm]{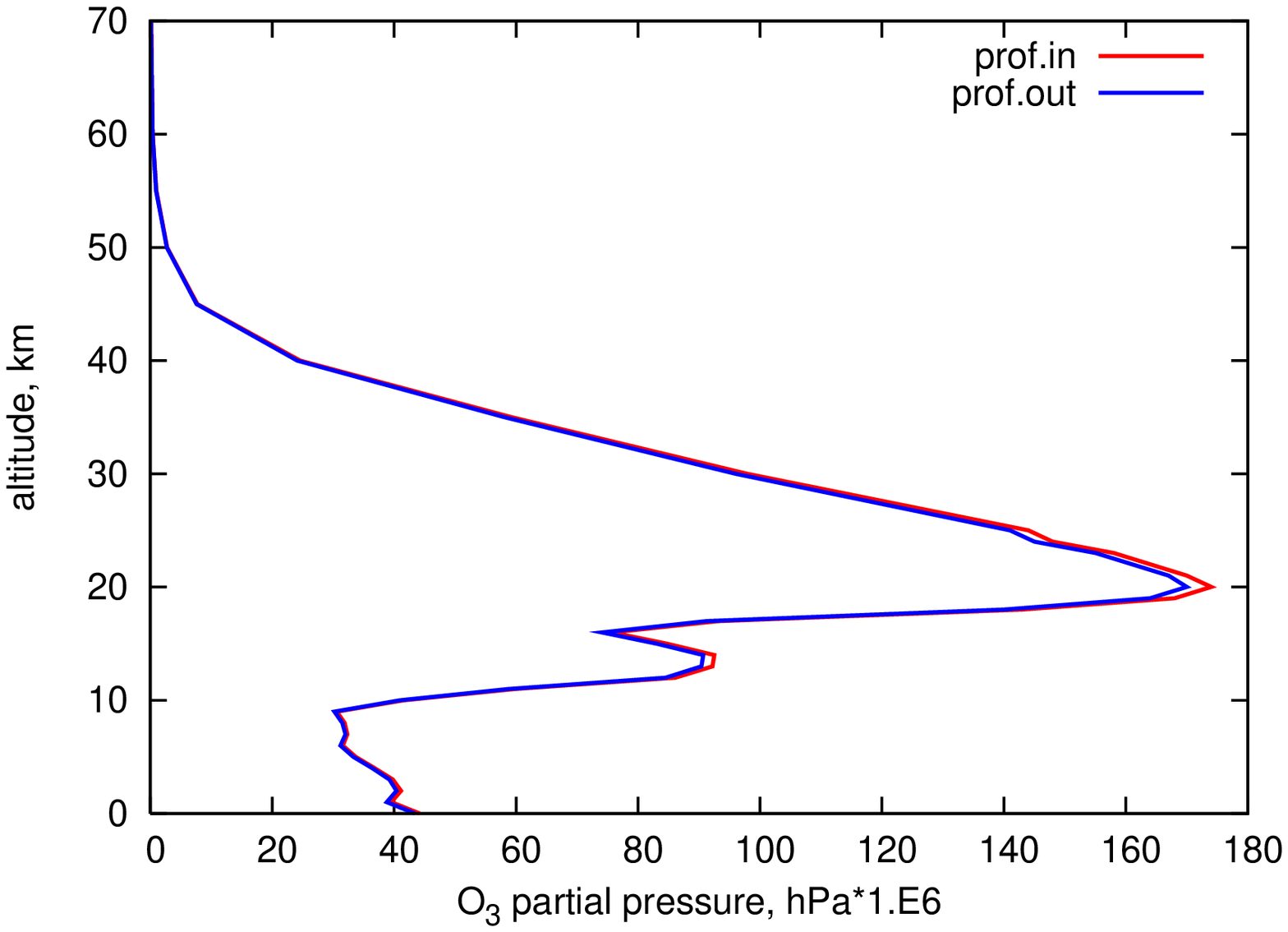}
\includegraphics[width=8cm]{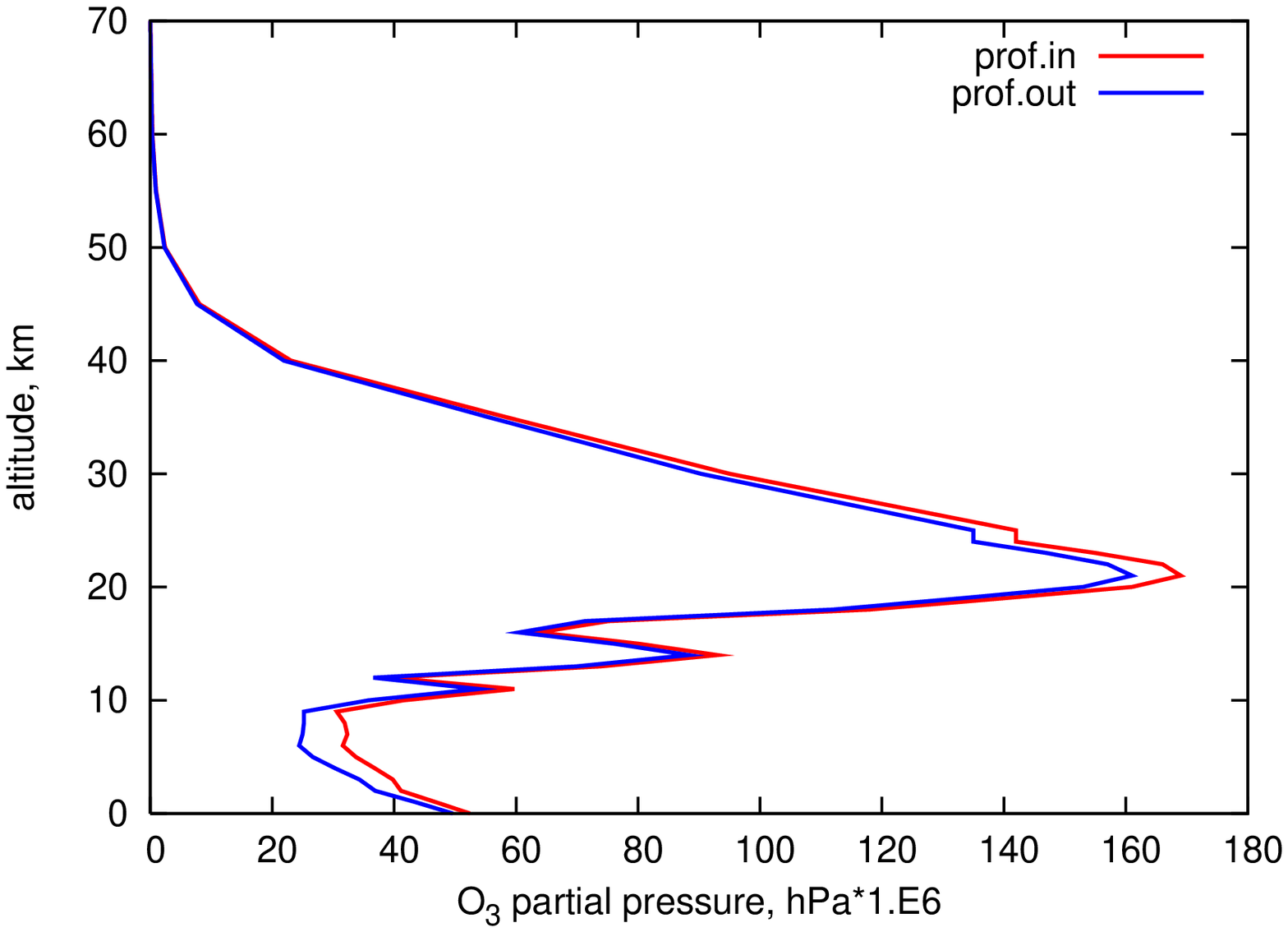}
\includegraphics[width=8cm]{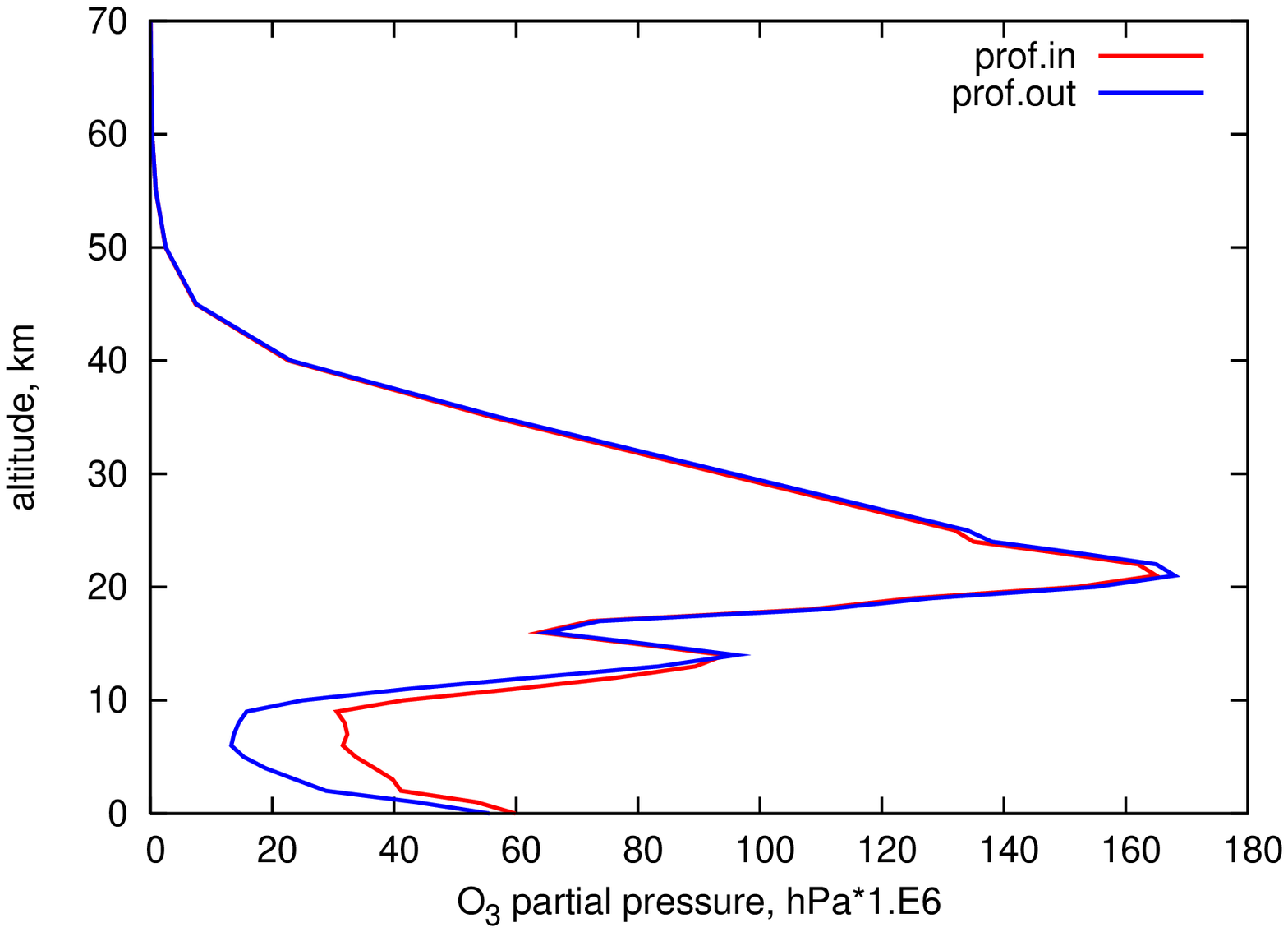}
      \caption{
The retrieved vertical atmospheric ozone profiles for a series of days
marked by a stratospheric intrusion, all recorded at the maximum solar elevation
on the 23rd, 25th, 26th 27th and the 28th of April 2007 from left to right. The 
red line denotes the a-priori input ozone profile for MODTRAN modeling, and the 
blue line denotes the outputted atmospheric ozone profile which gave the best 
model fit to the observed spectra around the 9.6 micron ozone band. In green 
the vertical ozone profile simultaneously recorded by the EOS-Aura 
TES instrument is shown, confirming the accurateness of the FTIR spectral fit.
              }
         \label{Fig5}
   \end{figure}

  \begin{figure}
  \centering
 \includegraphics[width=12cm]{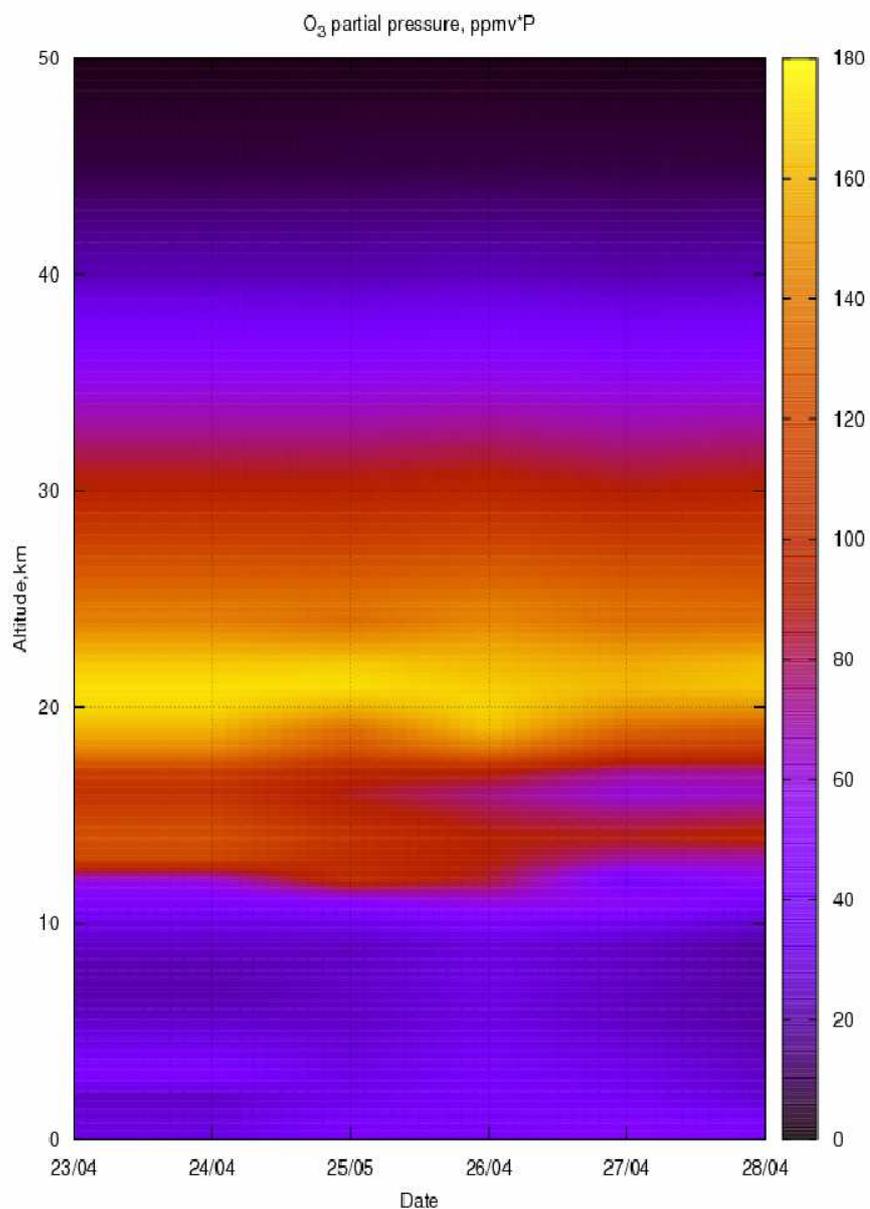}
      \caption{
 A spring episode in our data revealing a stratospheric intrusion 
occurring between April 23-28, 2007. On the left we see the intrusion of 
stratospheric ozone coming down into the troposphere, we see its sinking 
for 23-26 April and a further dissipation and sinking to surface layer of 
26-28 April 2007. Please note the high level of detail presented in this 
one-week plot revealing the vertical resolution and accuracy 
of our ground-based method that can be achieved on a continuous basis. 
              }
         \label{Fig6}
   \end{figure}

  \begin{figure}
  \centering
\includegraphics[width=8cm]{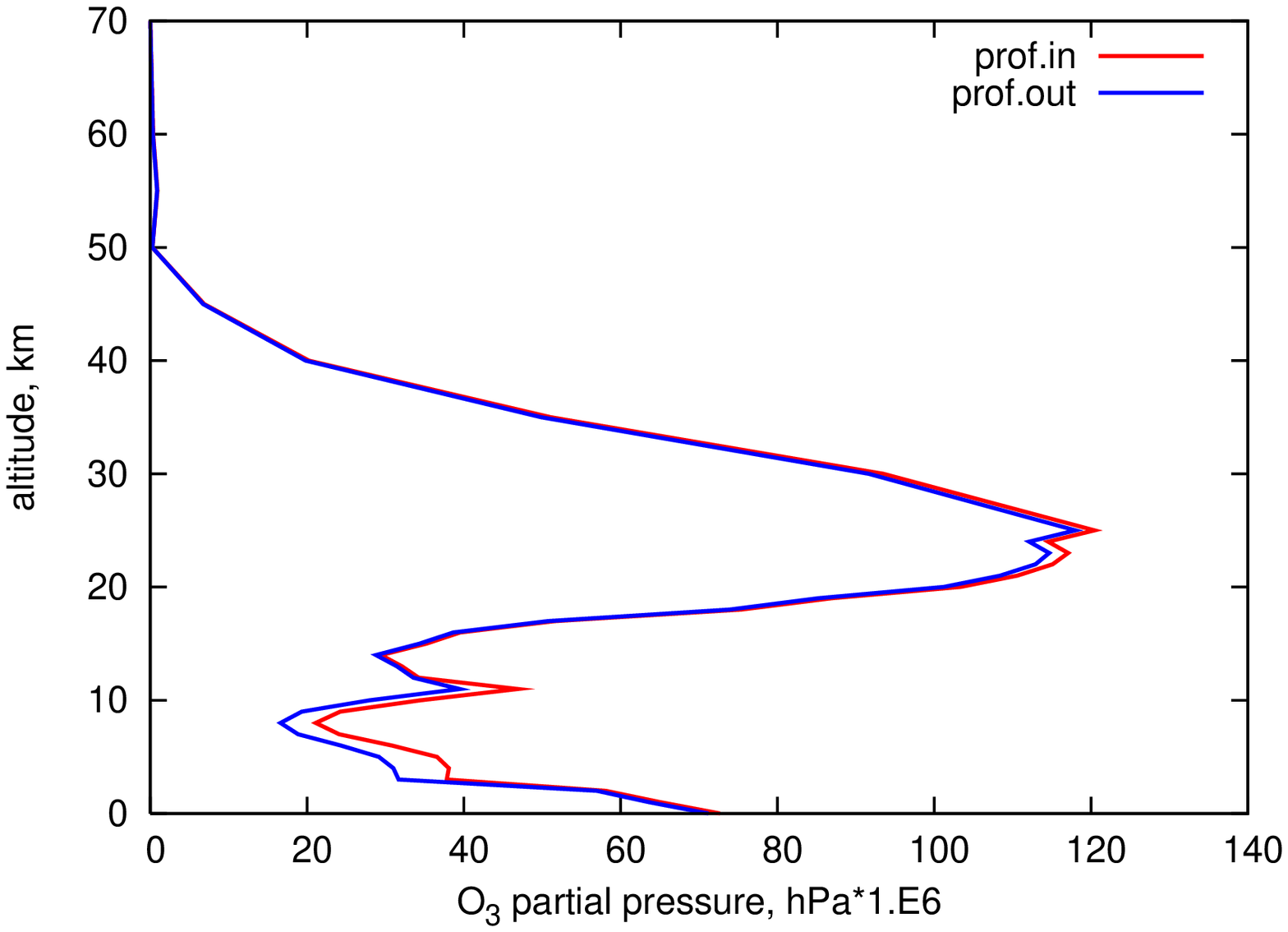}
\includegraphics[width=8cm]{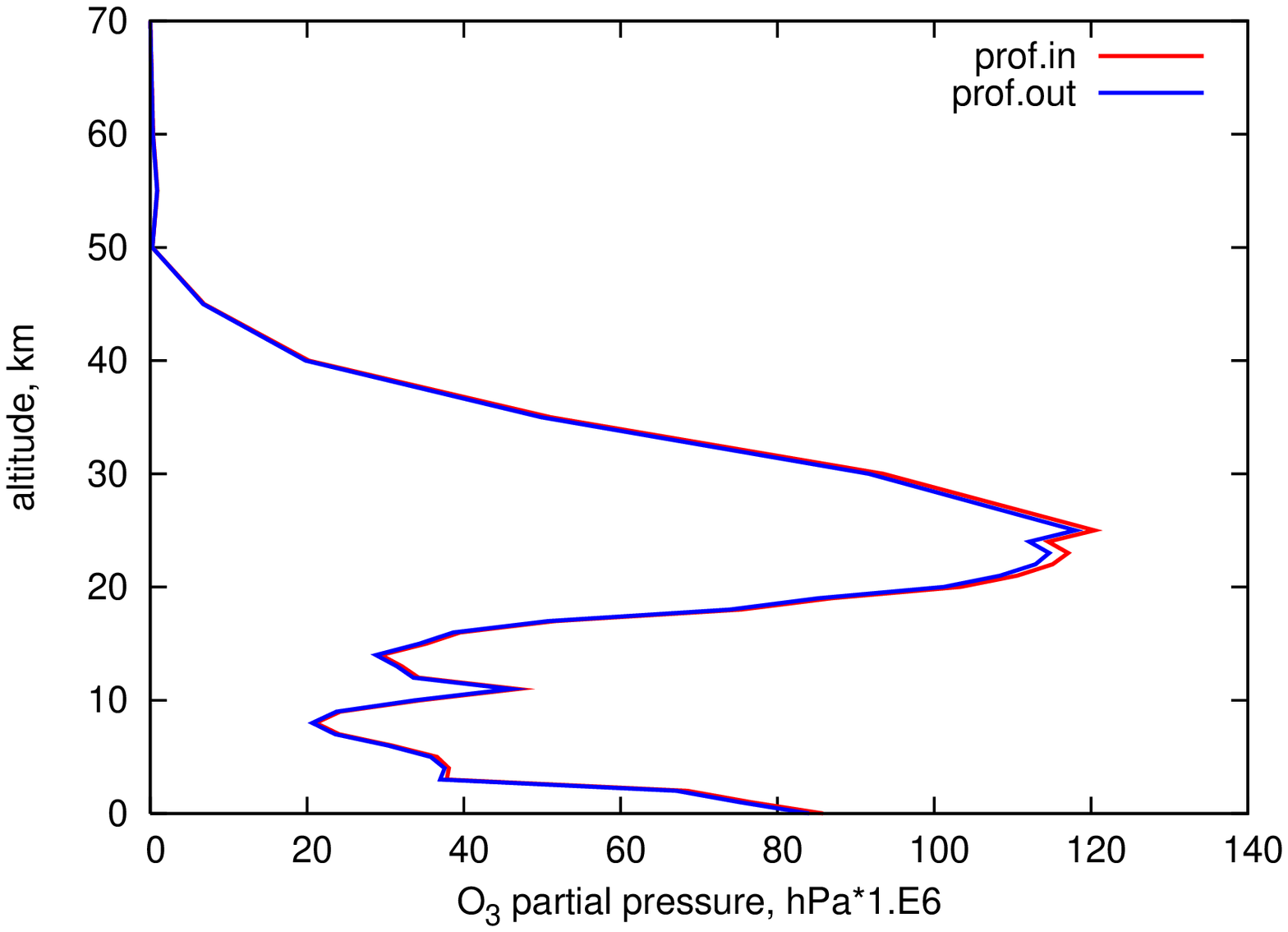}
\includegraphics[width=8cm]{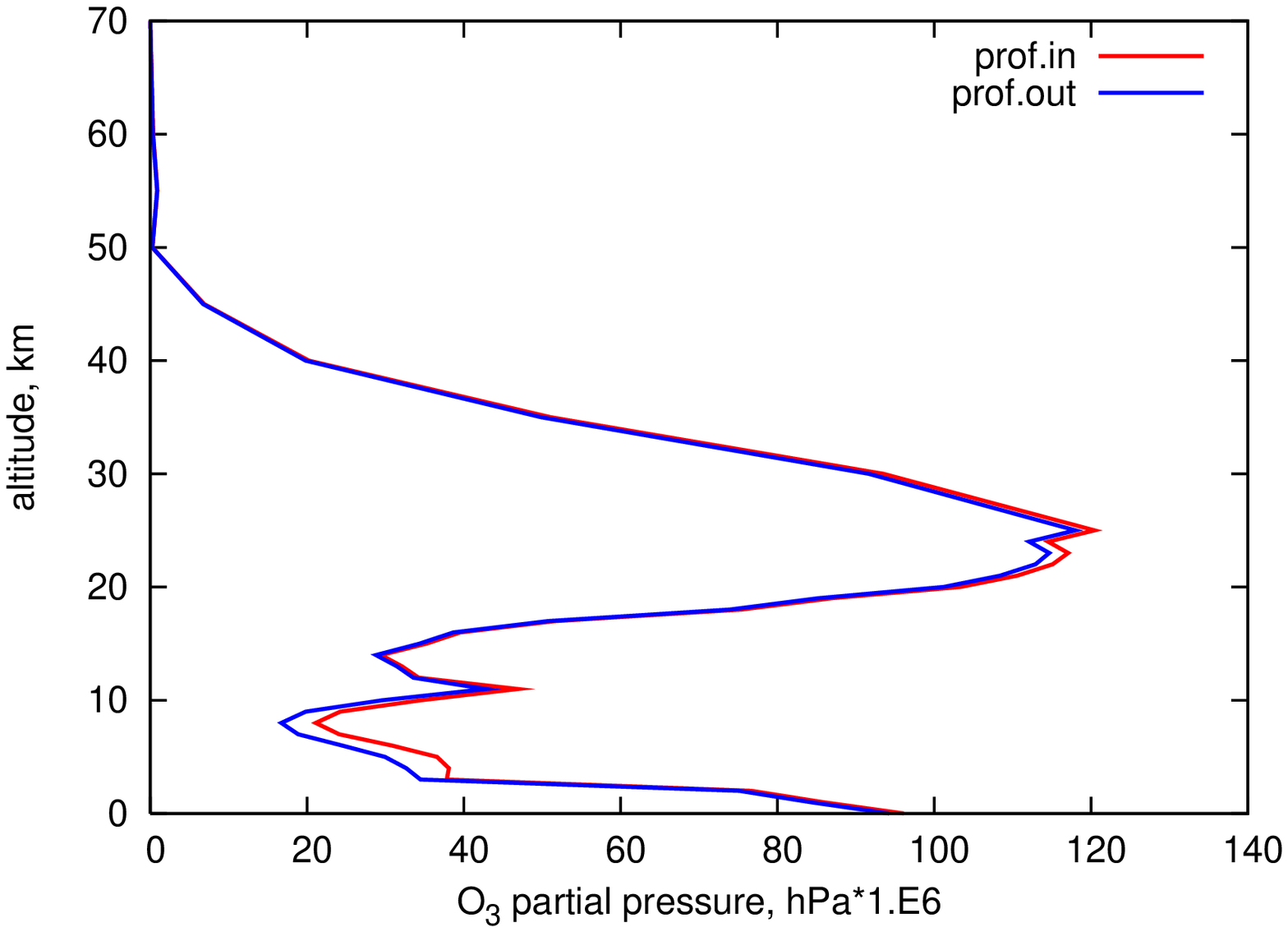}
\includegraphics[width=8cm]{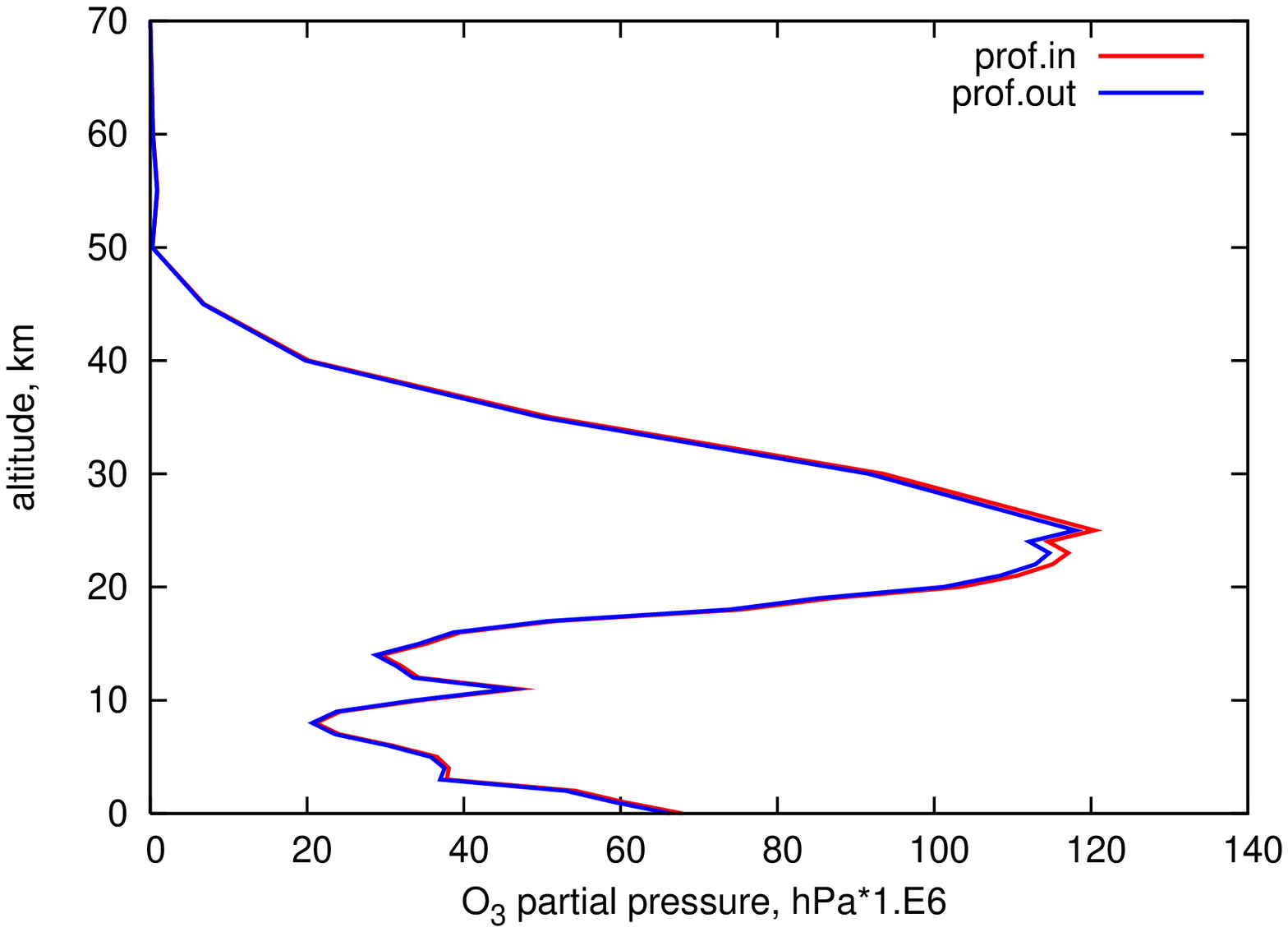}
\includegraphics[width=8cm]{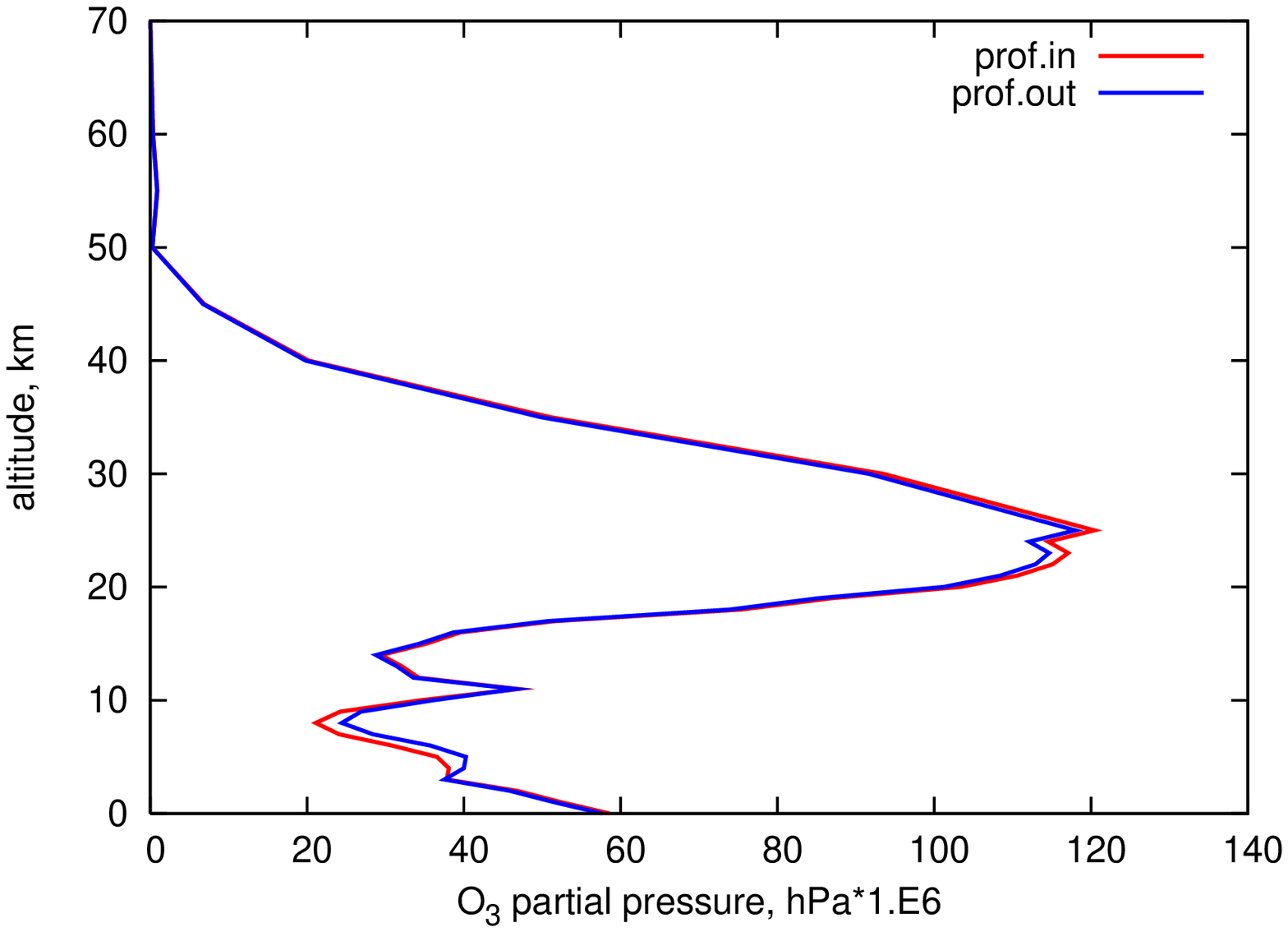}
\includegraphics[width=8cm]{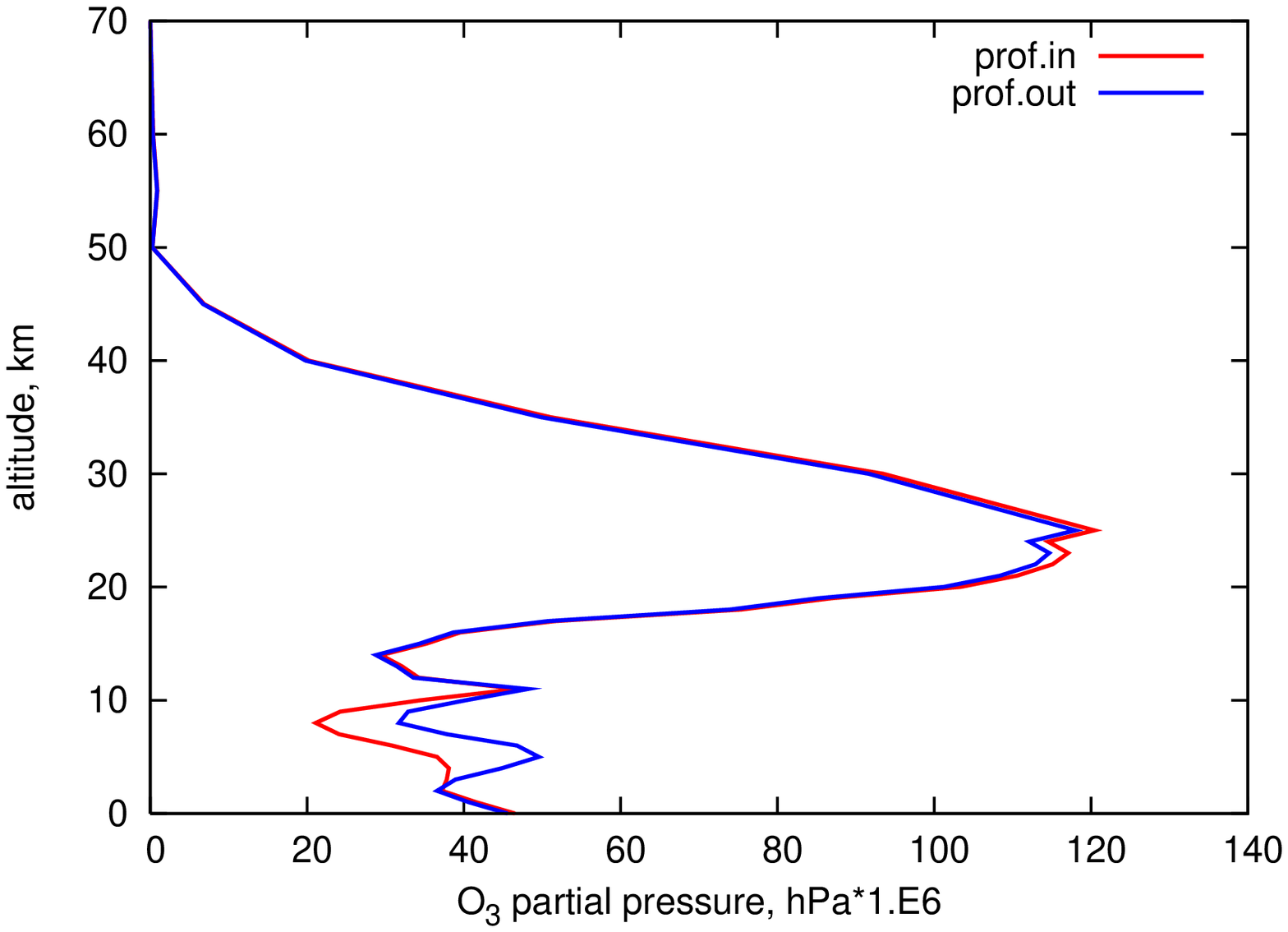}
      \caption{
The retrieved vertical atmospheric ozone profiles for the 18th of 
July 2007: From left to right we depicts the profiles for 13h35m, 14h52m, 
16h10m, 17h10m, 18h15m and 19h27m of local time respectively. 
The color coding of the profiles is the same as for Figure 5. 
              }
         \label{Fig7}
   \end{figure}

  \begin{figure}
  \centering
 \includegraphics[width=12cm]{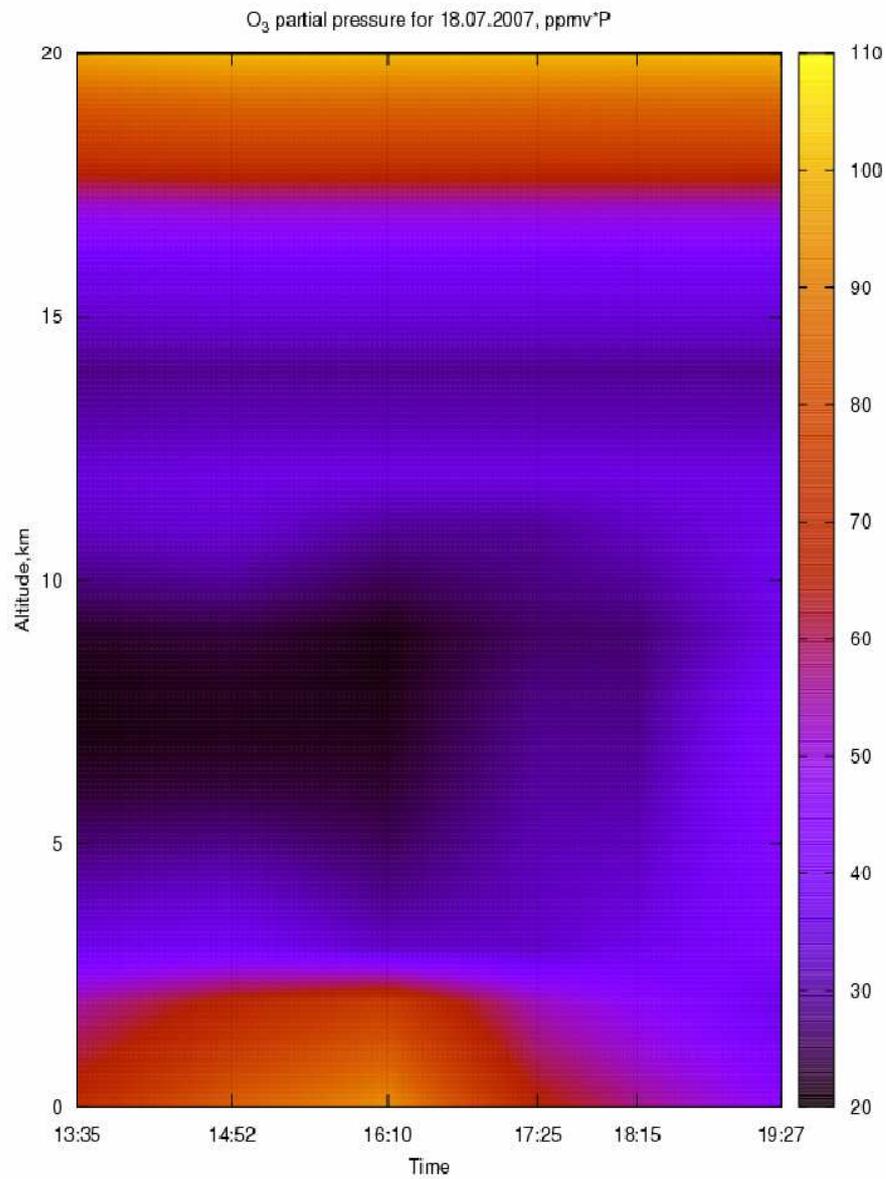}
      \caption{
A summer episode in our data revealing photochemical production of 
ozone in the surface layers of our polluted urban atmosphere at July 18, 2007 
(the time indicated on the bottom scale). Please note the high level of detail 
presented in this one day plot revealing the vertical resolution and 
accuracy of our ground-based method that can be achieved on a daily basis. 
              }
         \label{Fig8}
   \end{figure}

  \begin{figure}
  \centering
 \includegraphics[width=8cm]{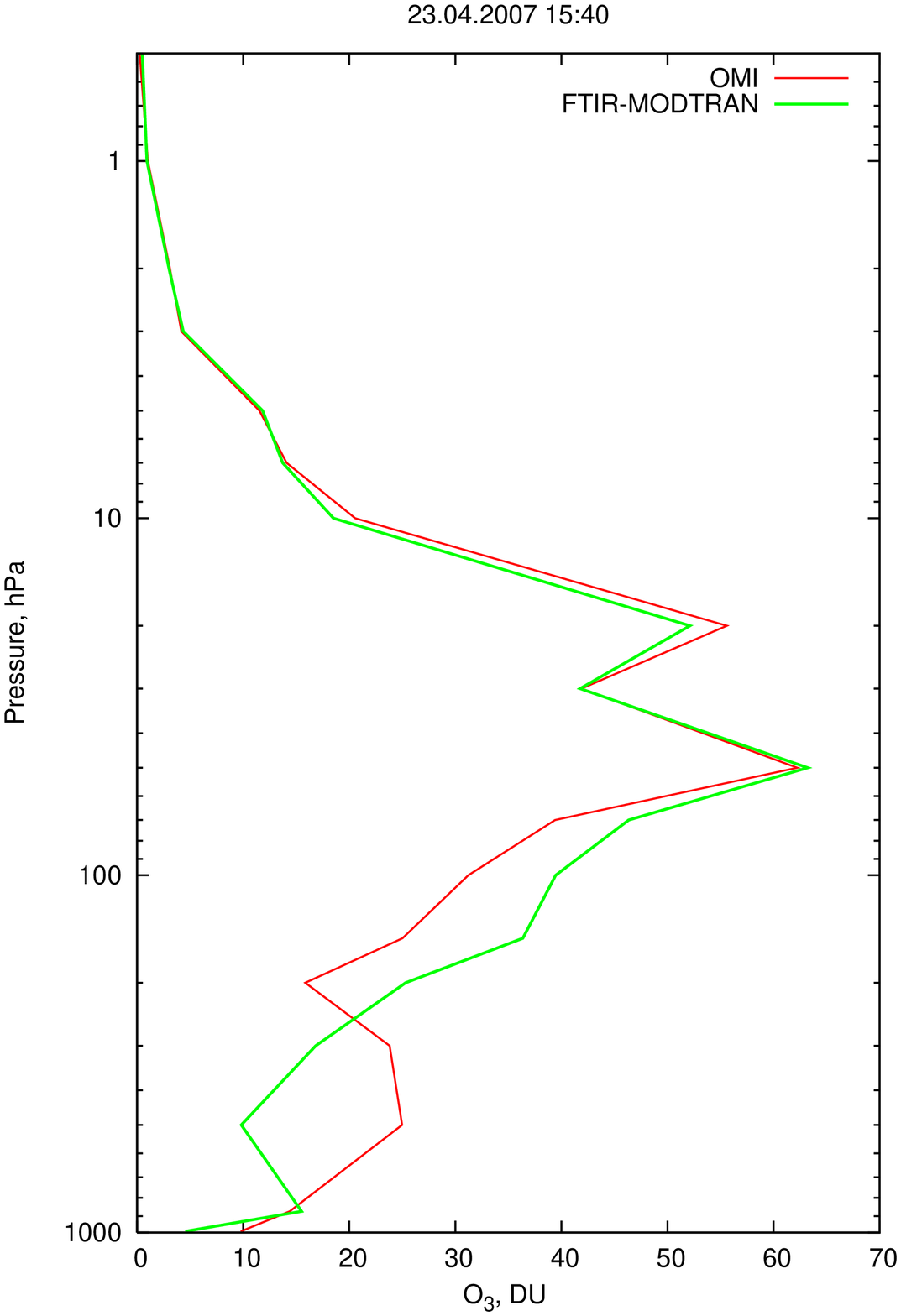}
 \includegraphics[width=8cm]{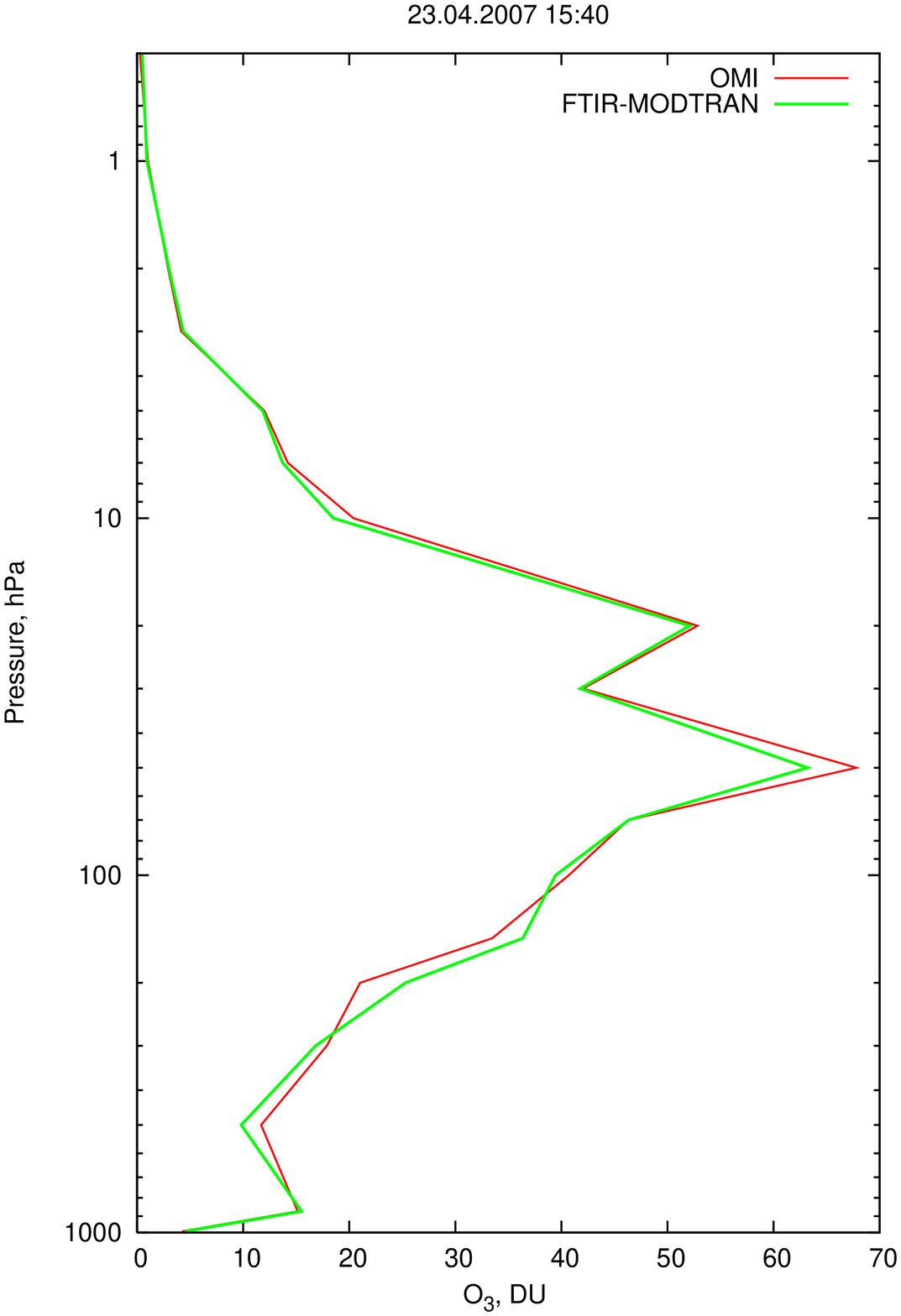}
      \caption{
A comparison of our retrieved ozone profiles (green line) with the 
OMI vertical ozone profiles (red line) for the day 23 April 2007. Ozone 
concentrations (vertical axis) are in Dobson Units per sub-columns for each of 
he 18 atmospheric layers that are used in the OMO3PR data.  In the left hand 
side image the red line corresponds to the OMI ozone profile averaged around 
Kiev ($8^o$ x $8^o$ of latitude and longitude) of the 2008 data version and the right
hand side image shows the same results for the newer 2009 OMI ozone profile 
data set. Overall but particularly in the troposphere the agreement is much 
better, where we note that for the ground-based data the surface
ozone concentrations are derived from and fixed by in-situ observations.
              }
         \label{Fig9}
   \end{figure}

%
\begin{table} 

\caption{The comparison of total ozone amounts data from satellite data 
OMI TOMS and OMI DOAS and ground-based FTIR observations (OMI TOMS --  FTIR / 
OMI DOAS -- FTIP)}

\vspace{0.5cm}
\small 
\begin{tabular}{|c|c|c|c|c|c|} 
\hline
Year&2005&2006&2007 &2007 &	2008\\
&&&(old OMI data)&(new OMI data)&\\
\hline
Mean&3.19 / 8.45&-0.25 / 0.37&-4.32 / -0.33&-7.54 / -6.63&-0.29 / 2.85\\
Standard Error&2.53 / 1.98&0.68 / 1.11&1.39 / 1.35&1.25 / 1.31&0.99 / 1.18\\
Median&1.45 / 10.70&0.115 / 0.16&-2.98 / -0.05&-7.59 / -4.09&0.57 / 3.51\\
Standard Deviation&13.41 / 10.50&5.37 / 8.77&10.88 / 10.66&9.00 / 9.06&7.52 / 8.88\\
Count&28/28&62 / 62&61 / 62&52 / 54&57 / 57\\
Slope&0.68 / 0.94&1.03 / 1.07&0.91 / 0.88&0.95 / 0.93	&0.93 / 0.88\\
Correlation&0.60 / 0.76&0.98 / 0.96&0.97 / 0.98&0.98 / 0.98	&0.97 / 0.97\\
R-squared&0.36 / 0.57&0.97 / 0.92&0.94 / 0.95&0.95 / 0.95&0.95 / 0.94\\
\hline 
\end{tabular} 
\end{table} 

\end{document}